\documentclass[12 pt, amsfonts, amssymb,color]{article}

\usepackage{amsmath,amssymb,amsfonts,latexsym,float,graphics,epsfig,xcolor}

\begin{document}

\renewcommand\theequation{\arabic{section}.\arabic{equation}}
\catcode`@=11 \@addtoreset{equation}{section}
\newtheorem{axiom}{Definition}[section]
\newtheorem{theorem}{Theorem}[section]
\newtheorem{axiom2}{Example}[section]
\newtheorem{lem}{Lemma}[section]
\newtheorem{prop}{Proposition}[section]
\newtheorem{cor}{Corollary}[section]
\newcommand{\be}{\begin{equation}}
\newcommand{\ee}{\end{equation}}

\newcommand{\equal}{\!\!\!&=&\!\!\!}
\newcommand{\rd}{\partial}
\newcommand{\g}{\hat {\cal G}}
\newcommand{\bo}{\bigodot}
\newcommand{\res}{\mathop{\mbox{\rm res}}}
\newcommand{\diag}{\mathop{\mbox{\rm diag}}}
\newcommand{\Tr}{\mathop{\mbox{\rm Tr}}}
\newcommand{\const}{\mbox{\rm const.}\;}
\newcommand{\cA}{{\cal A}}
\newcommand{\bA}{{\bf A}}
\newcommand{\Abar}{{\bar{A}}}
\newcommand{\cAbar}{{\bar{\cA}}}
\newcommand{\bAbar}{{\bar{\bA}}}
\newcommand{\cB}{{\cal B}}
\newcommand{\bB}{{\bf B}}
\newcommand{\Bbar}{{\bar{B}}}
\newcommand{\cBbar}{{\bar{\cB}}}
\newcommand{\bBbar}{{\bar{\bB}}}
\newcommand{\bC}{{\bf C}}
\newcommand{\cbar}{{\bar{c}}}
\newcommand{\Cbar}{{\bar{C}}}
\newcommand{\Hbar}{{\bar{H}}}
\newcommand{\cL}{{\cal L}}
\newcommand{\bL}{{\bf L}}
\newcommand{\Lbar}{{\bar{L}}}
\newcommand{\cLbar}{{\bar{\cL}}}
\newcommand{\bLbar}{{\bar{\bL}}}
\newcommand{\cM}{{\cal M}}
\newcommand{\bM}{{\bf M}}
\newcommand{\Mbar}{{\bar{M}}}
\newcommand{\cMbar}{{\bar{\cM}}}
\newcommand{\bMbar}{{\bar{\bM}}}
\newcommand{\cP}{{\cal P}}
\newcommand{\cQ}{{\cal Q}}
\newcommand{\bU}{{\bf U}}
\newcommand{\bR}{{\bf R}}
\newcommand{\cW}{{\cal W}}
\newcommand{\bW}{{\bf W}}
\newcommand{\bZ}{{\bf Z}}
\newcommand{\Wbar}{{\bar{W}}}
\newcommand{\Xbar}{{\bar{X}}}
\newcommand{\cWbar}{{\bar{\cW}}}
\newcommand{\bWbar}{{\bar{\bW}}}
\newcommand{\abar}{{\bar{a}}}
\newcommand{\nbar}{{\bar{n}}}
\newcommand{\pbar}{{\bar{p}}}
\newcommand{\tbar}{{\bar{t}}}
\newcommand{\ubar}{{\bar{u}}}
\newcommand{\utilde}{\tilde{u}}
\newcommand{\vbar}{{\bar{v}}}
\newcommand{\wbar}{{\bar{w}}}
\newcommand{\phibar}{{\bar{\phi}}}
\newcommand{\Psibar}{{\bar{\Psi}}}
\newcommand{\bLambda}{{\bf \Lambda}}
\newcommand{\bDelta}{{\bf \Delta}}
\newcommand{\p}{\partial}
\newcommand{\om}{{\Omega \cal G}}
\newcommand{\ID}{{\mathbb{D}}}
\newcommand{\pr}{{\prime}}
\newcommand{\prr}{{\prime\prime}}
\newcommand{\prrr}{{\prime\prime\prime}}
\title{ On the construction of almost general solutions for PDEs arising in nonlinear optics}
\author{A Ghose-Choudhury\footnote{E-mail: aghosechoudhury@gmail.com}
\and
Sudip Garai\footnote{E-mail: sudip.dhwu@gmail.com}\\
\\
Department of Physics, Diamond Harbour Women's University,\\ D. H. Road, Sarisha 743368, West Bengal, India
}

\bigskip

\date{\today}

\maketitle

\bigskip

\begin{abstract}
In this  communication we consider the widely used nonlinear  Fokas-Lenells equation, the cubic focussing nonlinear Schr\"{o}dinger equation in (2+1)-dimensions and the coupled  Drinfel'd-Sokolov-Wilson equation and attempt to construct almost general solutions for the envelope of the wave packet by means of the travelling wave ansatz. The obtained solutions have been expressed in terms of Jacobi elliptic sine function from which one can obtain the solitary wave (particular) solutions by imposing appropriate conditions on the roots of certain quartic polynomials.
\end{abstract}

\smallskip

\paragraph{Mathematics Classification (2010)}: 34C14, 34C20.

\smallskip


\section{Introduction}

In recent times nonlinear differential equations arising in context of optical fibers have attracted a lot of attention to the research community owing to their applicability in technological advancements of optical communication\cite{kudnew1,kudnew2,kudnew3,kudnew4,kudcha,kudop}. In these field the majority of the partial differential equations (PDEs) may be viewed as the generalizations of the fundamental nonlinear Schr\"{o}dinger equation. In most of the cases these equations are non-integrable in nature. Henceforth one usually searches for the particular solutions, especially the solitary wave solutions. In such PDEs, there are several methods in connection with obtaining the solitary wave solutions via travelling wave ansatz.\cite{Tanh0,K1,K2,K3,GG1,Exp1,AC,Hirota,CW,CAZ,sg10,sg11,kud4,sg12,sg13}. The method of obtaining the general solution of a nonlinear ODE is an important aspect of any mathematical analysis. Recently, Kudryashov has written extensively on this aspect of constructing the general solutions  for the equations describing the propagation of pulses through optical fibers\cite{KudMain}.

\smallskip

In this communication, we consider the Fokas-Lenells equation and the cubic focussing NLS equation \cite{FL1,FL2} which are widely used in nonlinear optics and also a coupled system called the Drinfel'd-Sokolov-Wilson equation and attempt to construct the general solution for the envelope of the wave packet by means of travelling waves.  In the following section we consider the Fokas-Lenells equation.

\section{General solutions of the reduced Fokas-Lenells equation}

The well known Fokas-Lenells equation is given by
\be\label{FL1} iq_t+a_1q_{xx}+a_2q_{xt}+(bq+i\sigma q_x)|q|^2 =i[\alpha q_x +\lambda (|q|^2q)_x +\mu(|q|^2)_x q].\ee

Here as elsewhere the subscripts denote partial derivatives and on the left hand side the coefficients are as follows: $a_1 \rightarrow $ group velocity dispersal, $a_2 \rightarrow$ spatio-temporal dispersal, $b \rightarrow$ self phase modulation \&  $\sigma \rightarrow$ nonlinear dispersal. On the right hand side the coefficients $\alpha$, $\lambda$ and $\mu$ represent respectively as: the intermodal, self steepening and nonlinear dispersal. The Fokas-Lenells equation essentially describes the propagation of the nonlinear optical pulses in monomode fibers when higher-order nonlinear effects have been taken into consideration. We look for the solutions of Eq.(\ref{FL1}) in the form of travelling wave ansatz, viz.
\be\label{FL2} q(x, t)=U(\eta)e^{i\phi(x, t)},\;\;\;\eta=x-vt,\;\;\;\phi(x, t)=-\kappa x-\omega t +\theta_0.\ee
 On substituting this form of the solution and separating the real and imaginary parts, we get the following equations:
 \be\label{FLreal}(a_1-va_2)U^{\prime\prime}+(b+\sigma \kappa -\lambda \kappa) U^3 +(a_2\omega \kappa -\omega -\alpha \kappa -a_1 \kappa^2)U=0,\ee
 \be\label{FLima}
 [a_2(v\kappa +\omega)-(v+2a_1\kappa)-\alpha]U^\prime +(\sigma-3\lambda-2\mu)U^2U^\prime=0.\ee
One way to make further progress with the above equations is to require that the coefficients of Eq.(\ref{FLima}) vanish, which serves to determine the parameter $v=(a_2\omega-\alpha-2a_1\kappa)/(1-a_2\kappa)$ and $\sigma=3\lambda+2\mu$. Clearly, we must have $a_2\kappa\ne 1$. Let us now write Eq.(\ref{FLreal}) as
 \be\label{FL3}AU^{\prime\prime}+BU^3 -CU=0,\ee where
 \be\label{Fl3a} A=(a_1-v a_2), \;\;\;B=b+(\sigma-\lambda\kappa), \;\;\;C=(\omega+a_1\kappa^2 -\omega \kappa a_2 +\alpha \kappa),\ee and the $^\prime$ denotes differentiation with respect to the variable $\eta$.
 Multiplying  Eq.(\ref{FL3}) with $U^\prime$ and integrating once leads to
 \be\label{FL4} U^{\prime 2} +\frac{B}{2A} U^4 -\frac{C}{A} U^2 +C_0=0,\ee where $C_0$ is an arbitrary constant of integration. In order to construct the  general solution of Eq.(\ref{FL4}) it is absolutely necessary that we retain this arbitrary constant of integration. We make a change of the variable $\eta$ to,  $\bar{x}=m\eta$, whence  Eq.(\ref{FL4}) becomes
 \be\label{FL5} U_{\bar{x}}^2 +\frac{M}{m^2}\left( U^4 -\frac{N}{M} U^2 +\frac{C_0}{M}\right)=0.\ee Here we have set, $M=B/2A$ and $N=C/A$, respectively. The roots of, $U^4-(N/M) U^2+(C_0/M)=0$, are  given by
 $\pm\sqrt{u_{\pm}}$, where
 $$u_{\pm}=\frac{1}{2}\left[\frac{N}{M} \pm \sqrt{\frac{N^2}{M^2}-\frac{4C_0}{M}}\right].$$ This enables us to rewrite  Eq.(\ref{FL5}) as
 \be\label{FL6} U_{\bar{x}}^2 +\frac{M}{m^2} (U-u_1)(U-u_2)(U-u_3)(U-u_4)=0,\ee with
  \be \label{FL7}u_1=\sqrt{u_+},\;\; u_2=-\sqrt{u_+},\;\; u_3=\sqrt{u_-},\;\;u_4=-\sqrt{u_-}.\ee
 Following the reference\cite{KudMain}, we assume that the solution of Eq.(\ref{FL6}) is of the form
 \be\label{FL8} U(\bar{x})=u_1+\frac{(u_2-u_1) e}{V^2 +e},\ee where $V(\bar{x})$ is a new function which has to be determined and $e$ is a constant. Simple calculations now show that
 $$U_{\bar{x}}^2=\frac{4(u_2-u_1)^2 e^2 V^2 V_{\bar{x}}^2}{(V^2 +e)^4},$$
 $$U(\bar{x})-u_1=\frac{(u_2-u_1)e}{V^2 +e}, \;\;U(\bar{x})-u_2=\frac{(u_1-u_2)V^2}{V^2 +e},$$
 $$U(\bar{x})-u_3=(u_1-u_3)\left[1-\frac{(u_1-u_2)e}{(u_1-u_3)(V^2+e)}\right],
 \;\;U(\bar{x}) -u_4=(u_1-u_4)\left[1-\frac{(u_1-u_2)e}{(u_1-u_4)(V^2+e)}\right].$$
  Insertion of the above expressions into (\ref{FL6}) gives
  \be \label{FL9} V_{\bar{x}}^2 -\frac{Me}{4m^2} (u_3-u_2)(u_4-u_2)\left[1-\frac{(u_1-u_3)V^2}{(u_3-u_2)e}\right]
  \left[1-\frac{(u_1-u_4)V^2}{(u_4-u_2) e}\right]=0.\ee
  Let us choose the constants $e$ and $m$ as follows:
  \be\label{FL10} e=\frac{(u_1-u_3)}{(u_3-u_2)},\;\;\;\;m^2=\frac{M}{4}(u_4-u_2)(u_1-u_3).\ee  As a consequence we have
  \be\label{FL11} V_{\bar{x}}^2-[1-V^2][1-p^2 V^2]=0,\ee
  where
  $$p^2=\frac{(u_1-u_4)(u_3-u_2)}{(u_4-u_2)(u_1-u_3)}=\left(\frac{\sqrt{u_+}+\sqrt{u_-}}{\sqrt{u_+}-\sqrt{u_-}}\right)^2.$$
  It is well known that the solution of (\ref{FL11}) is given by the Jacobi elliptic sine function  i.e.,
  $V(\bar{x}, p^2)=sn(\bar{x}-\bar{x}_0; p^2)$ so that
  \be V(m\eta; p^2)=sn(m(\eta-\eta_0); p^2)=sn\left(\frac{\sqrt{M}}{2}(\sqrt{u_+}-\sqrt{u_-})(\eta-\eta_0); p^2\right).\ee
  Finally from (\ref{FL8}) using (\ref{FL10}) we obtain the solution for the envelope as
  \be\label{FL12}
  U(\bar{x})=\frac{u_1(u_3-u_2)V^2+u_2(u_1-u_3)}{(u_3-u_2)V^2 +(u_1-u_3)}=\sqrt{u_+}\left[\frac{(\sqrt{u_+}+\sqrt{u_-})V^2-(\sqrt{u_+}-\sqrt{u_-})}
  {(\sqrt{u_+}+\sqrt{u_-})V^2+(\sqrt{u_+}-\sqrt{u_-})}\right].\ee

\section{Cubic focusing NLS equation}

Here we take the simplest case of a (2+1)-dimensional cubic focussing cubic nonlinear Schr\"{o}dinger equation (NLS)\cite{zhong2008,dai2010,wangnls2016} of the form
  \be\label{NLS1} iq_t +q_{xx}+\mu q_{yy} +\sigma |q|^2q=0.\ee
We restrict our analysis to the elliptic case $\mu=1$ and consider the travelling wave ansatz $q(x, y, t) =U(z) e^{i(k_1 x+k_2 y -\omega t)}$, with $z=x+y-vt$. On substituting this into Eq.(\ref{NLS1}) and equating the real and imaginary parts we obtain the following equations:
   \be\label{NLS2a}2U_{zz} +(\omega -k_1^2-k_2^2)U +\sigma U^3=0,\ee
  \be\label{NLS2b} (-v +2k_1+2k_2)U_z=0.\ee
  From the last equation, it follows that $v=2(k_1+k_2)$ since $y_z\ne 0$.  Multiplying Eq.(\ref{NLS2a}) by $U_z$ and integrating once we get;
  \be U_z^2 +\frac{1}{2}(\omega -k_1^2-k_2^2)U^2 +\frac{\sigma}{4}U^4 +C_0=0,\ee with $C_0$ being a constant of integration. As before we write this equation in the form
  \be\label{NLS3}U_z^2+\frac{\sigma}{4}\left(U^4-\frac{2}{\sigma}(k_1^2+k_2^2-\omega )U^2 +\frac{4}{\sigma}C_0\right)=0.\ee Introducing the change of independent variable $\bar{z}=mz$, we have
  \be\label{NLS4} U_{\bar{z}}^2+\frac{\sigma}{4m^2}(U-u_1)(U-u_2)(U-u_3)(U-u_4)=0,\ee where the roots $u_i$ of the fourth-degree polynomial in Eq.(\ref{NLS3}) are given by
  $u_{1,2}=\pm \sqrt{u_+}$ and $u_{3,4}=\pm\sqrt{u_-}$
  with $$u_{\pm}=\frac{1}{\sigma}\left[(k_1^2+k_2^2-\omega)\pm \sqrt{(k_1^2+k_2^2-\omega)^2-4\sigma C_0}\right].$$ Eq.(\ref{NLS4}) is exactly similar to Eq.(\ref{FL6}) and hence proceeding in the same manner as described in the previous section its general solution can be stated as follows:
  \be\label{NLS5}U(\bar{z})=\sqrt{u_+}\left[\frac{(\sqrt{u_+}+\sqrt{u_-})V^2-(\sqrt{u_+}-\sqrt{u_-})}
  {(\sqrt{u_+}+\sqrt{u_-})V^2+(\sqrt{u_+}-\sqrt{u_-})}\right].\ee with
  $$V(\bar{z}=mz; p^2)=sn(\frac{\sqrt{\sigma}}{4}(\sqrt{u_+}-\sqrt{u_-})(z-z_0); p^2)$$ and
  $$m^2=\frac{\sigma}{16}(u_4-u_2)(u_1-u_3), \;\;\;p^2=\left(\frac{\sqrt{u_+}+\sqrt{u_-}}{\sqrt{u_+}-\sqrt{u_-}}\right)^2.$$

\section{The Drinfel'd-Sokolov-Wilson Equation}

The Drinfel'd-Sokolov-Wilson (DSW) equation is a coupled system of the NPDEs given by\cite{dswe}
  \be\label{DSW1} u_t+pvv_x=0,\;\;\;v_t+qv_{xxx}+ruv_x+su_xv=0\ee with $p,q,r$ and $s$ are being nonzero parameters. The system was first proposed by Drinfel'd and Sokolov \cite{DS1,DS2} and independently discovered by Wilson \cite{Wil}. When $p=3$, $q=r=2$ and $s=1$ it turns out to be integrable. We look for traveling wave solutions of the form $$u(\xi)=u(x, t),\;\;v(\xi)=v(x, t),\;\;\;\xi=x+\omega t.$$ Upon inserting the travelling wave forms into Eq.(\ref{DSW1}) we obtain
  \be\label{DSW2a} \omega u^\prime +pv v^\prime=0,\ee
  \be \label{DSW2b}\omega v^\prime +q v^{\prime\prime\prime} +ruv^\prime +su^\prime v=0.\ee
  From the first of these ODEs we obtain  upon integration
  \be \label{DSW3a}u=\frac{C_0 -p\frac{v^2}{2}}{\omega}.\ee
   Upon using this expression for $u$ in (\ref{DSW2b}) we have after two integrations the following first-order nonlinear ODE, viz.
   \be\label{DSW3b}v^{\prime 2} +M v^4 -N v^2 +C_1 v+C_2=0,\ee where $C_1$ and $C_2$ are constants of integration and
   $$M=-\frac{p(r+2s)}{12\omega q},\;\;\;N=-\frac{\omega^2+rC_0}{q\omega}.$$
    As before, we let $\bar{x}=m\xi$ whence we can write (\ref{DSW3b}) as
   \be\label{DSW4b} v_{\bar{x}}^2 +\frac{M}{m^2}(v^4-\frac{N}{M} v^2 +\frac{C_1}{M} v+\frac{C_2}{M})=0.\ee If $v_i(i=1,...,4)$ denote the roots of, $(v^4-\frac{N}{M} v^2 +\frac{C_1}{M} v+\frac{C_2}{M})=0$, we can rewrite the last equation as
   \be\label{DSW5b} v_{\bar{x}}^2 +\frac{M}{m^2}(v-v_1)(v-v_2)(v-v_3)(v-v_4)=0,\ee with $\sum_{i=1}^4 v_i=0$. We assume that the solution of (\ref{DSW5b}) is expressible in the form
   \be\label{DSW6b} v(\bar{x})=v_1+\frac{(v_2-v_1)e}{V^2 +e},\ee where $V(\bar{x})$ is a function to be determined and $e$ is a constant. It follows from our earlier analysis that  with
   $$e=\frac{v_1-v_3}{v_3-v_2},\;\;\mbox{and}\;\;m^2=\frac{M}{4}(v_4-v_2)(v_1-v_3)$$ the  function $V(\bar{x})$ satisfies the equation
   \be\label{DSW7b} V_{\bar{x}}^2-[1-V^2][1-p^2 V^2]=0,\ee with $p^2=\frac{(v_1-v_4)(v_3-v_2)}{(v_4-v_2)(v_1-v_3)}$ and that the solution is given by
   \be v(\bar{x}; p^2)=sn (\bar{x}-\bar{x}_0; p^2),\ee or more explicitly
   \be V(m\xi; p^2)=sn\left(\frac{1}{2}\sqrt{M(v_4-v_2)(v_1-v_3)}(\xi-\xi_0); p^2\right).\ee The final solution for $v(\xi)$ is obtained from (\ref{DSW6b}) and turns out to be after simplification:
   \be v(\xi)=\frac{v_1(v_3-v_2)V^2+v_2(v_1-v_3)}{(v_3-v_2)V^2+(v_1-v_3)}.\ee Knowing $v(\xi)$ it is straightforward to obtain $u(\xi)$ from (\ref{DSW3a}).

\section{Discussion}

In this article we have attempted to highlight the fact that while using the travelling wave ansatz for dealing with nonlinear PDEs instead of arbitrarily setting the constants of integration, resulting from  reduction of order, to be zero  and thus finding only particular solutions; one can in many cases proceed towards a general solution by retaining the arbitrary constants of integration. In the cases presented in this article while constructing the general solution we have encountered quartic polynomials which can be factorized. The resulting solutions are all expressed in terms of the Jacobi elliptic sine function. It is clearly possible to obtain solitary wave (particular) solutions by imposing appropriate conditions on the roots. This has been explicitly discussed in the reference\cite{KudMain}. Of course it goes without saying that this procedure for  constructing the general solution fails when quintic or higher order polynomials are encountered. Lastly, we should mention that the general solutions we have discussed  are actually the general solutions of  `reduced equations' as there are in almost all cases certain conditions on the parameters involved in the equation.

\section*{Acknowledgments}

We thank \emph{Professor Nikolay Kudryashov} for drawing our attention to the new approach and for the encouragement.


\end{document}